\documentclass[12pt,nofootinbib,amsmath]{revtex4}
\usepackage{graphicx}

\begin{document}

\title{Non-Adiabatic Perturbations from Single-Field Inflation}

\author{C. Armend\'ariz-Pic\'on}
\affiliation{Enrico  Fermi Institute and  Department of  Astronomy and
  Astrophysics, \\University of Chicago.}

\begin{abstract}
  If the inflaton decays into several components during reheating, and
  if    the    corresponding   decay    rates    are   functions    of
  spacetime-dependent quantities,  it is possible  to generate entropy
  perturbations  after a  stage  of single-field  inflation.  In  this
  paper,  I  present  a  simple  toy  example  that  illustrates  this
  possibility.  In the  example, the decay rates of  the inflaton into
  ``matter'' and  ``radiation'' are  different functions of  the total
  energy  density.  In  particular cases,  one can  exactly  solve the
  equations  of motion both  for background  and perturbations  in the
  long-wavelength limit, and show that entropy perturbations do indeed
  arise.  Beyond  these specific examples, I attempt  to identify what
  are  the essential  ingredients  responsible for  the generation  of
  entropy  perturbations  after single-field  inflation,  and to  what
  extent  these  elements are  expected  to  be  present in  realistic
  models.
\end{abstract}

\maketitle

\section{Introduction}
Inflation is part of the modern cosmological standard paradigm because
some     inflationary    models     can    fit     the    observations
\cite{WMAP,SDSSWMAP}.   In particular,  it  is often  argued that  the
``simplest'' inflationary models \cite{inflation} yield a homogeneous,
isotropic,  spatially flat  universe  slightly perturbed  by a  nearly
scale invariant spectrum of Gaussian, adiabatic density perturbations.
However, one  has to be  cautious to generically ascribe  to inflation
the successes of only a particular  set of models.  In fact, there are
inflationary  scenarios  that   yield  anisotropic  \cite{Ford},  open
\cite{open} or closed \cite{closed} universes, strongly tilted spectra
\cite{power-law}   and,   in   multiple-field   models,   non-Gaussian
\cite{non-Gaussian}      or     non-adiabatic     \cite{non-adiabatic}
perturbations.  As it turns out, even single-field inflationary models
can    also    produce    significant    amounts    of    isocurvature
perturbations\footnote{I  use the words  ``entropy'', ``isocurvature''
  and ``non-adiabatic'' interchangeably.}.

The simplest inflationary models fall within the class of single-field
models.  In  the latter,  inflation is driven  by a single  field, the
inflaton, which is assumed to be the only relevant dynamical component
during  inflation  (in  addition   to  the  metric).   In  particular,
perturbations  are  imprinted  only  on  the  inflaton  field.   After
inflation, these  perturbations are transferred to  the decay products
of the inflaton during  reheating.  Reheating therefore is an integral
part of any inflationary model.  Without reheating, the universe would
end up  dominated by a  scalar field in  an otherwise empty  universe. 
The reheating  mechanism also affects the  predictions of inflationary
models, since  observable quantities, like the  spectral index, depend
on the reheating temperature.

It is believed that, regardless of the details of the reheating stage,
the density perturbations  seeded during single-field inflation remain
adiabatic.   Historically, the first  works \cite{BST}  that partially
justified this belief  assumed that the inflaton decays  into a single
component.  If the inflaton  completely decays into a single component
during  reheating,  only  that  component  emerges  and  survives  the
reheating  stage, so no  entropy perturbations  are possible.   If the
inflaton  decays   into  several   components,  it  has   been  argued
\cite{WMLL} that perturbations are  expected to be adiabatic, though I
am not aware of  a general rigorous proof of adiabaticity\footnote{See
  \textit{Note added in proof}.}.  In fact, such a proof would be quite
difficult to construct, as the couplings of the inflaton to matter and
the  inner  workings  of  the  reheating  mechanism  are  hardly  ever
specified or actually known.

The  purpose of  this paper  is  to show  that, indeed,  perturbations
seeded during a stage of  single-field inflation can be non-adiabatic. 
In  order  to  do  so,  it  will  be  enough  to  construct  a  single
counter-example.  The essential element  of the counter-example is the
non-universal   dependence  of   the   inflaton  decay   rates  on   a
spacetime-varying  quantity,  which happens  to  be  the total  energy
density in  this particular  case. This asymmetry  in the  decay rates
finally  results  in  the  generation of  entropy  perturbations  from
perturbations that are initially adiabatic.

\section{A Counter-Example}
Our  goal  is  to  describe  the  reheating  process  at  the  end  of
single-field  inflation  and  to  compute the  amplitudes  of  density
perturbations at  the end of  reheating.  The basics of  reheating are
reviewed in \cite{KoLiSt}, which focuses on background quantities.  In
order to  follow the evolution of  perturbations, additional machinery
is  needed.   Here,  I  will   rely  on  the  formalism  developed  in
\cite{MWU},  which  extends  the  work of  \cite{KodamaSasaki}.   This
formalism  has been  applied in  related contexts  by  several authors
\cite{MatarreseRiotto, MazumdarPostma, GMW}.

Suppose  that  the  inflationary  stage  is  driven  by  the  inflaton
$\varphi$.  After the end of inflation $\varphi$ oscillates around the
minimum of its potential and decays  into the particles it couples to. 
In order to describe the decay process it is convenient to resort to a
perfect fluid  description of the different  components involved.  Let
$p_\alpha$  and $\rho_\alpha$  generically denote  their  pressure and
energy   density.    These   fluids   might   include   the   inflaton
($\alpha=\varphi$),    radiation   ($\alpha=r$)    or    dark   matter
($\alpha=m$).   When  referring to  a  generic  component,  be it  the
inflaton  or its decay  components, I  use greek  subindices ($\alpha,
\beta,  \ldots$).   When referring  to  the  decay  components of  the
inflaton I use  latin lower-case ones ($a, b,  \ldots$).  The inflaton
is labeled by $\varphi$.

\subsection{Background}
After the end of a sufficiently long period of scalar-driven inflation
the universe is homogeneous, isotropic  and spatially flat.  In such a
universe, the  background equations of motion of  the different fluids
during the reheating process are
\begin{equation}\label{eq:bmotion}
  \dot{\rho}_\alpha+3H(1+w_\alpha)\rho_\alpha=\Gamma_\alpha\rho_\varphi,
\end{equation}  
where a dot  denotes a derivative with respect to  cosmic time, $H$ is
the Hubble constant, and $w_\alpha=p_\alpha/\rho_\alpha$.  The term on
the  right-hand side of  the equation  accounts for  the decay  of the
inflaton $\varphi$  into the different matter  fields.  In particular,
$-\Gamma_a$ is  the decay  rate of  the inflaton  into the
fluid  $a$.  Conservation  of  the total  energy-momentum tensor  then
implies
\begin{equation}\label{eq:bconservation}
  \sum_\alpha \Gamma_\alpha=0.
\end{equation} 
It   is  conventionally   assumed  that   the  (scalar)   decay  rates
$\Gamma_\alpha$  are constant.  Here,  we shall  be general  and allow
spacetime-dependent     ones     \cite{DGZ,    Kofman,MatarreseRiotto,
  MazumdarPostma,  BrandenbergerFinelli}.  In  single-field inflation,
spacetime-varying decay rates can arise for instance from a dependence
on  the  inflaton  $\varphi$,  or  on the  temperature  of  radiation,
$\Gamma_\alpha(T)$,   where   $T\sim\rho_r^{1/4}$  \cite{KNR}.    More
generally, one could consider decay rates that depend on the different
energy   densities,  $\Gamma_\alpha=\Gamma_\alpha(\rho_\beta)$,  which
keeps  the   system  of  Eqs.   (\ref{eq:bmotion})   closed.   In  the
following, I assume for simplicity that the decay rates only depend on
the total energy density, $\Gamma_\alpha=\Gamma_\alpha(\rho)$.  Though
this  last   assumption  is  not   an  essential  ingredient   of  the
counter-example,   it  will   considerably  simplify   the  analytical
treatment of the equations.

If the inflaton  potential $V$ around its minimum  $\varphi=0$ has the
form  $V\propto  \varphi^n$,  on  average,  the  oscillating  inflaton
behaves    as   a    perfect    fluid   with    equation   of    state
$w_\varphi=(n-2)/(n+2)$ \cite{Turner}.  For simplicity, I shall assume
that  all  the  fluids  have  the  same equation  of  state  $w$.   In
particular,  I  assume  that  the  inflaton has  the  same  (constant)
equation of state as its decay products $w=w_\alpha$.  This assumption
is also  realistic, as  we could consider  a quartic potential  for an
inflaton that decays into relativistic particles ($w=1/3$).

Summing  over all  components $\alpha$  in Eq.  (\ref{eq:bmotion}) and
using Eq.   (\ref{eq:bconservation}), I obtain the  equation of motion
for the total energy  density $\dot{\rho}+3H(1+w)\rho=0$, which can be
readily integrated,
\begin{equation}\label{eq:brho}
  \rho=\rho_0 \left(\frac{t_0}{t}\right)^2.
\end{equation}
Here, $\rho_0$ is the value of  the total energy density at an initial
time $t_0$, which is taken to  be right at the beginning of reheating. 
If the decay rates only depend  on $\rho$, then it is also possible to
integrate Eq.   (\ref{eq:bmotion}) for the inflaton  $\varphi$ and its
decay products $a$,
\begin{equation}\label{eq:rhointegrals}
  \rho_\varphi=\rho^0_\varphi \left(\frac{t_0}{t}\right)^2 
  \exp\left(\int_{t_0}^t \Gamma_\varphi(\rho)\, d\tilde{t}\right), \quad
  \rho_a=\left[\int_{t_0}^t \Gamma_a(\rho)\, \rho_\varphi\cdot
  \left(\frac{\tilde{t}}{t_0}\right)^2d\tilde{t}\right]
\left(\frac{t_0}{t}\right)^2.
\end{equation}
In  the previous equations  I have  assumed that  there are  no matter
fields at  time $t_0$, $\rho_a^0=0$, so  that $\rho_\varphi^0=\rho_0$. 
Because  $\rho(t)$  is known,  these  equations  provide closed  exact
solutions    to   the    equations   of    motion   for    any   given
$\Gamma_\alpha(\rho)$.  For definiteness, let  me at this point assume
a  particular functional  dependence of  $\Gamma_\alpha$.   Expand the
decay rates in  a series in inverse cosmic time,  and neglect terms of
quadratic or higher order.   Because the $\Gamma_\alpha$ depend on the
total energy density, this translates into
\begin{equation}\label{eq:gamma-t}
  \Gamma_\alpha(\rho)=\Gamma_\alpha^0+\Gamma_\alpha^1 
  \left(\frac{\rho}{\rho_0}\right)^{1/2},
\end{equation}
where  $\Gamma_\alpha^0$   and  $\Gamma_\alpha^1$  are   two  constant
coefficients, and  $\rho_0$ is again  the total energy density  at the
initial   time  $t_0$.    At   any   rate,  it   follows   from  Eq.   
(\ref{eq:bconservation}) that
\begin{equation}
  \sum_\alpha \Gamma_\alpha^0=\sum_\alpha \Gamma_\alpha^1=0.
\end{equation}
The functional  form (\ref{eq:gamma-t}) allows  me to write  the final
energy densities in closed form.  Plugging Eq. (\ref{eq:gamma-t}) into
Eqs.   (\ref{eq:rhointegrals}) I  find in  the limit  of  large cosmic
times
\begin{equation}
  \rho_\varphi=\rho_0
  \left(\frac{t}{t_0}\right)^{\Gamma_\varphi^1 t_0-2}
  \cdot\exp\left[\Gamma_\varphi^0(t-t_0)\right]
\end{equation}
and
\begin{equation}\label{eq:rhoagamma}
 \rho_a=\rho_0\cdot D \cdot \left[
    \Gamma_a^1 t_0\cdot
   \Upsilon(\Gamma_\varphi^1\, t_0, -\Gamma_\varphi^0\, t_0,)
   -\frac{\Gamma_a^0}{\Gamma_\varphi^0}
   \Upsilon(\Gamma_\varphi^1\, t_0+1, -\Gamma_\varphi^0\,t_0)\right]
 \left(\frac{t_0}{t}\right)^2.
\end{equation} 
The (dimensionless) constant $D$ is given by
\begin{equation}
  D=\left(-\Gamma_\varphi^0t_0\right)^{-\Gamma_\varphi^1t_0}
  \cdot e^{-\Gamma_\varphi^0 t_0},
\end{equation}
and  $\Upsilon(x,y)$   is  the  incomplete   Gamma  function\footnote{
  $\Upsilon(x,y)\equiv \int_y^\infty  t^{x-1}\,e^{-t}dt$.  I apologize
  for the  idiosyncratic notation.}.   Because the inflaton  decays, I
assume  that $\Gamma_\varphi^0$ and  $\Gamma_\varphi^1$ are  negative. 
At  late  times  the  evolution  of the  inflaton  energy  density  is
determined  by  the  exponential  suppression,  which  rapidly  drives
$\rho_\varphi$ to zero.  Also at  late times the matter energy density
is proportional to $t^{-2}$, as expected.

\subsection{Perturbations}
In spatially flat gauge, the perturbed metric has the form
\begin{equation}
  ds^2=-(1+2\phi)dt^2+2 a B_{,i}\, dt dx^i+
  a^2 \delta_{ij} \,dx^i dx^j,
\end{equation}
where $\phi$ and $B$ are scalar metric perturbations. This gauge turns
to  be convenient  because  in the  long-wavelength  limit, where  one
expects  spatial  divergence  of  the  three-momentum  and  the  shear
gradient to  be negligible  \cite{MWU}, the perturbation  equations do
not explicitly contain metric perturbations,
\begin{equation}\label{eq:pmotion}
  \delta\dot{\rho}_\alpha+3H(1+w)\delta\rho_\alpha
  =-\frac{1}{2}\Gamma_\alpha \rho_\varphi 
  \frac{\delta\rho}{\rho}+\delta\Gamma_\alpha\rho_\varphi
  +\Gamma_\alpha\delta\rho_\varphi.   
\end{equation}
To arrive  at this expression,  I have substituted  Eq. (37) into  Eq. 
(32) of  reference  \cite{MWU}  (note  that in  spatially  flat  gauge
$\psi\equiv0$).  Summing over all components in Eq. (\ref{eq:pmotion})
and      using     Eq.      (\ref{eq:bconservation})      I     obtain
$\delta\dot{\rho}+3H(1+w)\delta\rho=0$.   Hence,  the   total  density
contrast is constant,
\begin{equation}\label{eq:contrast}
  \frac{\delta{\rho}}{\rho}=\frac{\delta{\rho}_0}{\rho_0}.
\end{equation}
Note that  $\delta\rho/\rho$ being constant does not  imply that there
are no entropy  perturbations.  Entropy perturbations among components
with the  same equation of  state (the case  we consider here)  do not
source  changes in  $\delta\rho/\rho$.  Below  I will  comment  on how
entropy perturbations among components with the same equation of state
might induce changes in  $\delta\rho/\rho$ and thus have observational
effects.

Because  the  evolution of  the  background  and $\delta\rho/\rho$  is
explicitly  known,  Eq.  (\ref{eq:pmotion})  can  also be  immediately
integrated,
\begin{equation}\label{eq:drhophiintegral}
  \delta\rho_\varphi=\delta\rho_0
  \left[1+\int_{t_0}^t \left(\frac{d\Gamma_\varphi}{d\rho}
      \rho-\frac{1}{2}
      \Gamma_\varphi\right)d\tilde{t}\right]
  \left(\frac{t_0}{t}\right)^2
      \exp\left(\int_{t_0}^t \Gamma_\varphi\, d\tilde{t}\right),
\end{equation}
and
\begin{equation}\label{eq:drhoaintegral}
  \delta\rho_a=
  \left[\int_{t_0}^t \left(\frac{\delta\rho_0}{\rho_0}\frac{d\Gamma_a}{d\rho}
      \rho\,\rho_\varphi-
      \frac{1}{2}\frac{\delta\rho_0}{\rho_0}\Gamma_a\, \rho_\varphi+
      \Gamma_a \delta\rho_\varphi\right)
    \left(\frac{\tilde{t}}{t_0}\right)^2 d\tilde{t}\right]
  \left(\frac{t_0}{t}\right)^2,
\end{equation}
where I have assumed that there are no matter perturbations initially,
$\delta\rho_a^0=0$,     which    by     the    way     also    implies
$\delta\rho_0=\delta\rho_\varphi^0$.        Substituting      Eq.      
(\ref{eq:gamma-t})    into   Eqs.     (\ref{eq:drhophiintegral})   and
(\ref{eq:drhoaintegral}) I get after a bit of straightforward algebra
\begin{equation}
  \delta\rho_\varphi=\delta\rho_0\cdot
  \left[1-\frac{1}{2}\Gamma_\varphi^0 (t-t_0)\right]
  \cdot\left(\frac{t}{t_0}\right)^{\Gamma_\varphi^1 t_0-2}
  \cdot\exp\left[\Gamma_\varphi^0 (t-t_0)\right] 
\end{equation}
and
\begin{eqnarray}\label{eq:drhoagamma}
  \delta\rho_a=\delta\rho_0 \cdot D\cdot
  \Bigg[\frac{1}{2}\bigg(\Gamma_a^1t_0-\Gamma_a^0t_0-
    \lefteqn{\frac{\Gamma_a^0}{\Gamma_\varphi^0}\bigg)
  \Upsilon(1+\Gamma_\varphi^1t_0,-\Gamma_\varphi^0t_0)-
  \frac{1}{2}\frac{\Gamma_a^0}{\Gamma_\varphi^0}
  \Upsilon(2+\Gamma_\varphi^1t_0,-\Gamma_\varphi^0t_0)+}\nonumber \\
  &{} &+\left(\Gamma_a^1t_0+
    \frac{1}{2}\Gamma_a^1t_0 \Gamma_\varphi^0t_0\right)
  \Upsilon(\Gamma_\varphi^1t_0,-\Gamma_\varphi^0t_0)\Bigg]
  \left(\frac{t_0}{t}\right)^2.
\end{eqnarray}
Note that  the $\delta\rho_a$ only depend on  the $\Gamma_a^i$ through
the dimensionless quantity $\Gamma_a^i t_0\sim\Gamma_a^i/H_0$.  Recall
that within the  limit of late times, the  only approximation that has
been made so far is the  neglect of spatial gradients in the equations
of motion (which should be a good approximation in the long-wavelength
limit) \cite{MWU}.
 
\subsection{Observables}
A set of convenient quantities that characterize the perturbations are
the  variables  $\zeta$   and  $\zeta_\alpha$  \cite{Lyth,MWU},  which
describe   curvature  perturbations  in   spatial  slices   where  the
corresponding energy density is constant,
\begin{equation}\label{eq:zetaalpha}
    \zeta\equiv -H\frac{\delta\rho}{\dot{\rho}}
    =\frac{1}{3}\frac{1}{1+w}\frac{\delta\rho}{\rho},\quad 
    \zeta_\varphi\equiv -H\frac{\delta\rho_\varphi}{\dot{\rho_\varphi}}
    \approx-\frac{H}{\Gamma_\varphi}
    \frac{\delta\rho_\varphi}{\rho_\varphi}\quad  
    \text{and}\quad
    \zeta_a\equiv-H\frac{\delta\rho_a}{\dot{\rho}_a}
    \approx\frac{1}{3}\frac{1}{1+w}\frac{\delta\rho_a}{\rho_a}.
\end{equation}
The approximations in the last equations apply at late times, when the
inflaton  has already decayed.   Entropy perturbations  arise whenever
the $\zeta_\alpha$ are different for different fluid species.  A gauge
invariant quantity that characterizes such entropy perturbations is
\begin{equation}\label{eq:entropy}
  \mathcal{S}_{\alpha\beta}=3(\zeta_\alpha-\zeta_\beta).
\end{equation}
Entropy  perturbations are  important because  they source  changes in
$\zeta$.  Assuming that the perfect fluids have a constant equation of
state parameter, the $\zeta$ equation of motion is  
\begin{equation}
  \dot{\zeta}=\frac{H}{2\,\dot{\rho}^2}
  \sum_{\alpha\beta}\dot{\rho}_\alpha\dot{\rho}_\beta
  \cdot(w_\alpha-w_\beta)\cdot\mathcal{S}_{\alpha\beta}.
\end{equation}
Therefore,  in  the absence  of  entropy  perturbations  $\zeta$ is  a
conserved quantity (recall that we  only deal here with large scales). 
This conservation law has a  counterpart for the individual $\zeta_a$. 
It can  be also shown \cite{MWU}  that once the  inflaton has decayed,
$\zeta_a$  remains constant  if  the  fluid $a$  is  isentropic, i.e.  
$p_a=p_a(\rho_a)$.  Therefore, after inflaton decay $\mathcal{S}_{ab}$
is constant on large scales, regardless of how $\zeta$ evolves.

Current experimental results are consistent with a primordial spectrum
of   adiabatic  perturbations   \cite{SDSSWMAP},   though  significant
isocurvature components are still allowed \cite{CGBLR}. By definition,
perturbations  are  non-adiabatic  if  $\mathcal{S}_{\alpha\beta}$  is
nonzero for  any pair  of fluids. Suppose  momentarily that  the decay
rates are of the form
\begin{equation}\label{eq:cadiabatic}
  \Gamma_\alpha=\Gamma_\alpha^0\cdot f(\rho),
\end{equation}
where $\Gamma_\alpha^0$ is a constant and $f(\rho)$ is any function of
the total energy density that is common for all species.  Because both
$\delta\rho_a$  and  $\rho_a$  in Eqs.   (\ref{eq:drhoaintegral})  and
(\ref{eq:rhointegrals})   are  proportional   to   $\Gamma_a^0$,  this
constant cancels  in the  ratio $\zeta_a$, Eq.   (\ref{eq:zetaalpha}). 
Consequently, there are no entropy perturbations in the matter sector,
$\mathcal{S}_{ab}=0$.  In principle, entropy perturbations could arise
in  the  inflaton-matter  sector,  $\mathcal{S}_{\varphi  a}\neq  0$.  
However, if the inflaton decays completely, both its background energy
density and  perturbations vanish,  so they do  not contribute  to the
matter   budget   of   the   universe  after   reheating;   primordial
perturbations  might then  be effectively  regarded as  adiabatic.  An
alternative  interesting---albeit somewhat remote---possibility  is an
inflaton that does  not completely decay, say, because  its decay rate
drops  to zero sufficiently  fast.  This  frustrated decay  might then
lead to an inflaton which at late times comes to dominate the universe
as a  dark matter  form for  instance.  Anyway, to  summarize: if  Eq. 
(\ref{eq:cadiabatic})   holds  and   the  inflaton   entirely  decays,
perturbations     are     still     adiabatic     after     reheating,
$\mathcal{S}_{ab}=0$.

\begin{figure}
  \begin{center}
    \includegraphics{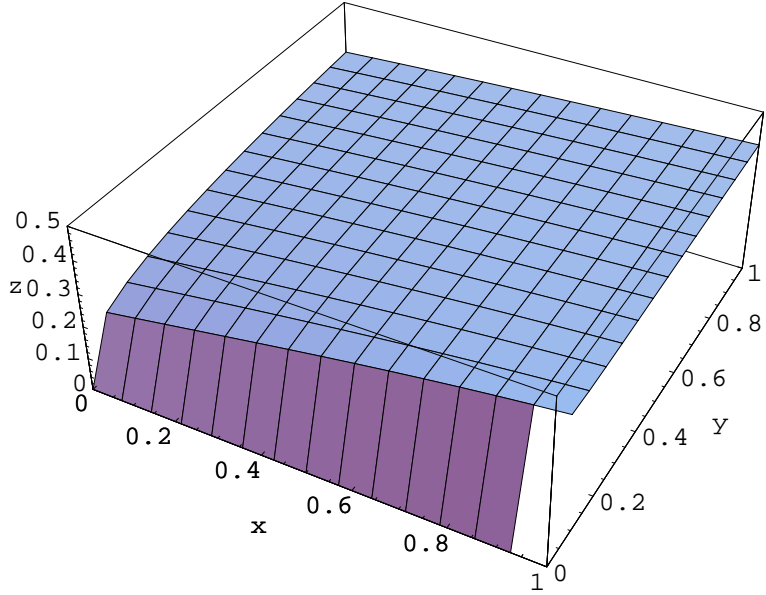}
    \caption{A plot of  the  relative  entropy
      perturbation  $\mathcal{S}_{mr}/(3\,  \zeta)$  ($z$-axis)  as  a
      function    of     $\Gamma_m^1\cdot    t_0$    ($x$-axis)    and
      $\Gamma_r^0\cdot  t_0$ ($y$-axis).   For  a large  set of  decay
      rates,  isocurvature perturbations  are  significant. Note  that
      $\Gamma_m^1\cdot   t_0=0$   or   $\Gamma_r^0\cdot   t_0=0$   are
      ill-defined limits (there is no decay into matter or radiation).
      \label{fig:relative}}
  \end{center}
\end{figure}

If  there is  more than  one component  involved in  the decay  of the
inflaton, say, dark matter ($m$) and radiation ($r$), and if the decay
rates  are  not constants,  there  is  no  reason to  expect  relation
(\ref{eq:cadiabatic})  to hold.  In  that case  one \emph{generically}
expects non-adiabatic perturbations. The source of these non-adiabatic
perturbations is the asymmetry in the evolution of the different decay
rates. To illustrate  the issue, let me consider  the following simple
case:
\begin{equation}\label{eq:example}
  \Gamma_\varphi^0=-\Gamma_r^0,\,  \Gamma_\varphi^1=-\Gamma_m^1,
  \quad\quad
  \Gamma_r^0> 0,\, \Gamma_r^1=0
  \quad\quad \text{and} \quad\quad 
  \Gamma_m^0=0,\,   \Gamma_m^1>0.
\end{equation}
Inserting Eqs.  (\ref{eq:example}) into Eqs.  (\ref{eq:rhoagamma}) and
(\ref{eq:drhoagamma}), and  using the definitions (\ref{eq:zetaalpha})
and  (\ref{eq:entropy})  I  find  that  the  radiation-matter  entropy
perturbation is given by
\begin{equation}
  \mathcal{S}_{mr}=\frac{1}{2}\frac{1}{1+w}\frac{\delta\rho_0}{\rho_0}
  \left[1 +
      \frac{\Upsilon(1-\Gamma_m^1t_0, \Gamma_r^0t_0)}
    {\Upsilon(-\Gamma_m^1t_0, \Gamma_r^0t_0)} -
    \frac{\Upsilon(2-\Gamma_m^1t_0, \Gamma_r^0t_0)}
    {\Upsilon(1-\Gamma_m^1t_0, \Gamma_r^0t_0)}\right].
\end{equation}
Fig.   \ref{fig:relative}  shows  the  relative  entropy  perturbation
$\mathcal{S}_{mr}/(3\, \zeta)$ as  a function of $\Gamma_r^0\,t_0$ and
$\Gamma_m^1t_0$.  As clearly seen in the Fig.  \ref{fig:relative}, the
relative entropy perturbation is  significant (i.e.  of order one) for
almost  any  set  of  decay   rates.   In  particular,  in  the  limit
$\Gamma_r^0  t_0,  \Gamma_m^1   t_0\gg  1  $,  ${\mathcal{S}_{mr}/(3\,
  \zeta)\approx  1/2}$.   The  sudden  drop of  the  relative  entropy
perturbation around  $\Gamma_r^0t_0=0$ is due to the  breakdown of our
results  for  that  value  (there  is no  exponential  suppression  in
integrals like (\ref{eq:rhointegrals})).

A feature of this concrete example  that I expect to survive in models
with less assumptions  is the linear dependence of  the matter density
perturbations  on  the initial  density  perturbation $\delta\rho_0$.  
Hence, if entropy perturbations  originate from reheating after single
field  inflation,  the  spectral  indices  of  entropy  and  curvature
perturbations  should be  identical, and  entropy and  curvature modes
should be also completely correlated.  Though this seems plausible and
might simply follow from linearity of the equations, as I mentioned at
the  beginning one should  be cautious  to generalize  statements only
known to apply in particular cases.

Although our example  does indeed contain non-adiabatic perturbations,
the Bardeen  variable $\zeta$  remains constant on  large scales,  Eq. 
(\ref{eq:contrast}). This happens because all components have the same
equation  of state.   The last  feature though  was introduced  in the
counter-example  merely to simplify  the equations,  and it  is likely
that   if  the  assumption   ${w_\alpha=w_\beta}$  is   dropped  while
maintaining  the  non-universal time-dependence  of  the decay  rates,
isocurvature  perturbations are  still  going to  be generated  during
reheating.

But even  if we  maintain the assumption  of common equation  of state
during reheating,  it is  possible to generate  changes in  $\zeta$ on
large  scales. Suppose  for  instance that  the  inflaton decays  into
radiation  ($r$)  and a  sufficiently  light  species  of dark  matter
particles ($m$).   Though dark matter  might behave as  a relativistic
component  during reheating ($w_m\approx  1/3$), once  its temperature
later drops below  its mass, it will start  behaving as a pressureless
component  $w_m\approx 0$.  The  non-vanishing (and  constant) entropy
perturbation  $\mathcal{S}_{mr}$  will  then, since  ${w_m\neq  w_r}$,
source  changes in  $\zeta$ on  large  scales and  have a  significant
impact on structure formation \cite{Wayne}.

\section{Conclusion}
In  order to  show that  single-field inflation  does  not generically
produce  a spectrum of  adiabatic perturbations,  it is  sufficient to
present a single  counter-example.  In this paper, I  have described a
toy reheating process that yields significant amounts of non-adiabatic
perturbations  after  single-field  inflation.  The  example  contains
several  ingredients.  Some of  them them  are superfluous,  and their
sole purpose is  to simplify the analysis.  Others  are essential, and
their presence alone is  likely to produce non-adiabatic perturbations
during   reheating.    The    essential   ingredient   here   is   the
``non-universal''  spacetime  dependence of  the  rates  at which  the
inflaton  decays  into   different  components.   Obviously,  such  an
asymmetric dependence can only occur  if the inflaton decays into more
than one fluid (radiation and dark matter for instance).  As a caveat,
let  me mention  that  I  have neglected  spatial  divergences of  the
momentum and  the shear  gradient in the  equations of motion  for the
perturbations.   These terms  are  expected to  be  negligible in  the
long-wavelength  limit,  though in  order  to  rigorously asses  their
importance,  one should  deal with  equations of  motion  containing a
single variable \cite{Mukhanov}.

Is such a non-universal behavior  of the decay rates expected to occur
in realistic models?  The universe contains several constituents, like
baryons, dark matter, photons, neutrinos and quintessence, so there is
no  reason for  the inflaton  to decay  into a  single  component.  In
addition,   reheating  proceeds   in  different   stages.   Initially,
parametric amplification is responsible for an explosive production of
particles known as preheating and subsequently, the inflaton decays by
conventional   (and  essentially  different)   perturbative  processes
\cite{KoLiSt}.  During  this last  stage of reheating  particles might
also  acquire thermal masses,  which again  render the  inflaton decay
rate  spacetime-dependent  \cite{KNR}.   Therefore,  it  is  extremely
plausible  that the  inflaton  decay rate  could  depend on  spacetime
quantities \cite{MatarreseRiotto}.  On top of that, the decay products
of  the  inflaton  might  be  very different,  like  dark  matter  and
radiation are, so there is no reason to assume that their couplings to
the  inflaton  and,  hence,  their corresponding  creation  rates  are
related to each other.  Overall, it  might very well be that the decay
rates  evolve in  a non-universal  way. In  that case,  rather  than a
spectrum of adiabatic perturbations,  the signature that points to the
single-field origin of the entropy perturbations in our example is the
complete correlation  of entropy and  curvature modes, and  the common
spectral index.

The counter-example  presented here  does not imply  that single-field
inflation  is  in  conflict  with  observations.   First,  significant
amounts of isocurvature are  still allowed by experiment \cite{CGBLR}. 
Second, there  are models  where perturbations still  remain adiabatic
after  reheating. In  fact,  the  mechanism I  have  described can  be
regarded  from a  different perspective.   Little is  known  about the
physics  of  reheating.   Hence,  the  possibility  that  isocurvature
perturbations  can  be efficiently  produced  during reheating  might,
together with  observational constraints, shed a fair  amount of light
into our understanding  and modeling of the reheating  process after a
stage of single-field inflation.

\section{Note added in proof}

After this  preprint was posted  on the archive,  S.  Weinberg and  S. 
Bashinsky  kindly  pointed  out   to  me  that,  under  quite  general
assumptions, the adiabaticity of  the perturbations generated during a
stage of single-field inflation  follows from the results presented in
\cite{Weinberg1}  and \cite{BashinskySeljak}.   As suggested  by them,
the origin  of the non-adiabatic  perturbations discussed here  is the
discontinuous change  of the decay  rate along a  spatial hypersurface
where the  total energy density is perturbed  \cite{Weinberg2}. Such a
change makes the perturbations  non-adiabatic already at the beginning
of  reheating.  The  nature of  the perturbations  generated  during a
stage   of  single-field   inflation  is   specifically   analyzed  in
\cite{Weinberg2}.

\begin{acknowledgments}
  It is a  pleasure to thank Sean Carroll, Rocky  Kolb, Eugene Lim and
  Slava  Mukhanov  for useful  comments  and  remarks.  I'm  specially
  indebted to  Chris Gordon,  Wayne Hu and  Lam Hui for  extensive and
  challenging  discussions.   Finally, I  want  to  thank also  Sergei
  Bashinsky and  Steven Weinberg for  clarifying correspondence.  This
  work has been supported by the US DOE grant DE-FG02-90ER40560.
\end{acknowledgments}

\end{document}